\documentclass[useAMS,usenatbib,usegraphicx]{mn2e}
\usepackage{amsmath}

\title{Sistematic error mitigation in multiple field astrometry }

\author[M. Gai]{M. Gai$^{1}$ 
\thanks { E-mail: gai@oato.inaf.it } 
\\ 
$^{1}$Istituto Nazionale di Astrofisica - Osservatorio Astronomico di 
Torino, V. Osservatorio 20, 10025 Pino T.se (TO), Italy }

\begin{document}

\date{}

\pagerange{\pageref{firstpage}--\pageref{lastpage}} \pubyear{2011}

\maketitle

\label{firstpage}

\begin{abstract}
Combination of more than two fields provides constraints on the 
systematic error of simultaneous observations. 
The concept is investigated in the context of the Gravitation 
Astrometric Measurement Experiment (GAME), which aims at measurement 
of the PPN parameter $\gamma$ at the $10^{-7}-10^{-8}$ level. 
Robust self-calibration and control of systematic error is crucial 
to the achievement of the precision goal. 
The present work is focused on the concept investigation and practical 
implementation strategy of systematic error control over four 
simultaneously observed fields, implementing a ``double differential'' 
measurement technique. 
Some basic requirements on geometry, observing and calibration strategy 
are derived, discussing the fundamental characteristics of the proposed 
concept. 
\end{abstract}


\section*{Introduction \label{sec:Introduction}}
The recent and ongoing global astrometry experiments implemented by
the European Space Agency, Hipparcos \citep{1997ESASP.402....1P} and 
Gaia \citep{2010HiA....15..816P,2010HiA....15..716S},
both take advantage of differential measurement on superposed fields,
using different beam combination concepts to materialise the {\em base 
angle} (BA) separating the two lines of sight (corresponding to the
directions on the sky matched to the centre, or a convenient reference
point, of the focal plane). 
In such instruments, the BA is a hardware defined parameter 
which must be stable to a high degree by construction, and whose 
actual value and secular evolution can be deduced and checked for 
consistency in the data reduction. 
\\ 
Some of the key concepts from Hipparcos and Gaia have been adopted by 
the Gravitation Astrometric Measurement Experiment (GAME) in 
order to translate high precision differential measurements on 
the focal plane of a telescope into high accuracy determination of 
relevant physical quantities 
\citep{2010HiA....15..325V}. 

In the following of the current section, the basic framework of GAME
is briefly recalled; then, in Sec.~\ref{sec:MultipleField}, the
potential benefit of multiplexing additional fields for simultaneous
observation through the same instrument is discussed, evidencing the
significant relaxation of BA requirements due to intrinsic error compensation.
In Sec.~\ref{sec:Implementation}, the optical implementation principle
is described, in terms of two cases of instrument size. In Sec.~\ref{sec:Calibration},
some relevant aspects of instrument self-calibration and monitoring
are reviewed. Finally, the main conclusions are drawn. 

The goal of GAME is the estimation of the $\gamma$ and $\beta$ 
parameters of the Parametrised Post-Newtonian (PPN) formulation of 
Einstein's General Relativity (GR) and competing gravitation theories. 
The GAME concept has been previously presented in the form of a small 
mission focused on deflection measurement with the name Gamma 
Astrometric Measurement Experiment, and recently submitted 
in an upgraded form and with the current name to the recent ESA call 
for Medium class mission within the Cosmic 
Vision 2015-2025 science programme. 
The main design driver of GAME is still $\gamma$, whereas $\beta$
and other astrophysical subjects are important additional 
science topics. 
\\ 
The parameters $\gamma$ and $\beta$ are related respectively to
the amount of curvature produced by mass, and to nonlinearity in the
superposition law of gravity \citep{2001lrevr...2..4..W}. 
The experiment of Dyson, Eddington and Davidson \citep{1920PTRSA.220..291D} 
gave the first confirmation of Einstein's General Relativity theory 
by observations of known stellar fields during the May 29th, 1919 eclipse 
on the island of Pr{\'\i}ncipe. 
It measured the apparent positions of a few stars during the eclipse, 
within a few degrees from the solar limb, compared to their unperturbed 
relative positions (e.g. in night time observations a few months away). 
The arc variation is interpreted in terms of light deflection, providing 
an estimate of the $\gamma$ parameter with precision limited to 10\%,
i.e. to a $10^{-1}$ accuracy estimation of the PPN $\gamma$ parameter.
The current best estimate of $\gamma$, from the Cassini experiment,
is in the $2\times10^{-5}$ range \citep{2003Natur.425..374B}, based
on the radio frequency shift technique, but with a significant burden
on calibration: the refraction in the average solar corona is estimated
to be of order of 0".1, at a radio wavelength of 20~cm,
for a pencil beam passing at a minimum distance of 4 solar radii from
the Sun centre \citep{1996SoPh..164...97G}, i.e. comparable with the 
angular value of light deflection on the same trajectory. 
The parameter $\beta$ was estimated by the classical GR test on 
\textit{perihelion precession}, by high precision astrometric measurement 
of Mercury orbit, and, more recently, by ``grand fit'' of a large set 
of observations of several Solar System objects 
\citep{2009IAU...261.0602L}. 

GAME will measure at micro-arcsec (hereafter, $\mu as$) level the
two-dimensional coordinates of stars in selected fields at $2^{\circ}$
from the Sun centre, hence their relative distance, by means of the
Hipparcos/Gaia beam combination principle and an optimised optical
scheme aimed at efficient rejection of the Solar photon flux and control
of instrumental effects. The same fields are observed at different
epochs, i.e. close to the Sun, in high deflection conditions, and
when the Sun has moved away by a significant amount along the Ecliptic,
i.e. with low deflection. 
The distance variation provides the $\gamma$ estimate, with photon 
limited precision in the $10^{-8}$ range, in a modern rendition of the 
1919 measurement. 
An experiment located in space is able to
overcome the limitations of the Dyson et al. experiment, due to the
{\em short eclipse duration}, the {\em high background flux}
from the solar corona, the {\em atmospheric disturbances} and the
{\em limited number of bright sources} accessible in a given eclipse.
\begin{figure}
\begin{centering}
\includegraphics[scale=0.55]{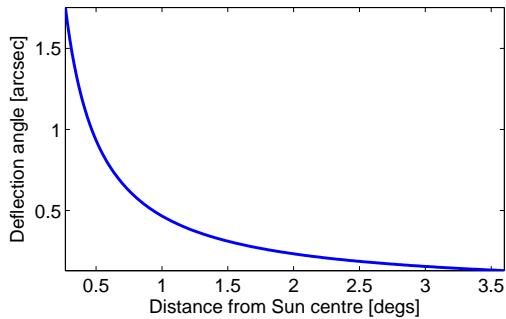}
\par\end{centering}
\caption{\label{fig:PlotDefl}Deflection $\Delta\psi$ vs. 
angular distance $\psi$ to the Sun centre. }
\end{figure}

The amount of deflection $\Delta\psi$ (Fig.~\ref{fig:PlotDefl})
affecting the photons from a star having geometric angular separation
$\psi$ with respect to the current position of an object with mass
$M$, as seen by an observer at a distance $d$, can be expressed
by the simplified formula \citep{1973grav.book.....M}  
\begin{equation}
\Delta\psi = 
\left(1+\gamma\right)\frac{GM}{c^{2}d}\sqrt{\frac{1+\cos\psi}{1-\cos\psi}}\,,
\label{eq:Misner}
\end{equation}
where $G$ is Newton's gravitation constant, $c$ the speed of light
and $\gamma$ Eddington's parameter. 
For a satellite in low Earth orbit, $d\simeq1\, AU$. 
Therefore, the PPN parameter $\gamma$ can be estimated by the measurement 
of the deflection angle, reversing Eq.~\ref{eq:Misner}: 
\begin{equation}
\gamma = 
\Delta\psi \cdot \frac{c^{2}d}{GM} \sqrt{\frac{1-\cos\psi}{1+\cos\psi}} - 1 
\,.
\label{eq:gamma}
\end{equation}

The GAME implementation has been considered at two levels of complexity
and performance, roughly corresponding to small and medium class ESA
space mission. In the former case, the telescope diameter is 0.75~m,
with focal length 25~m and field of view $14'\times14'$ 
\citep{2009SPIE.7438E..20G}. 
In the latter case, the instrument configuration is increased 
to telescope diameter 1.5~m, focal length 35~m and field of view
$30'\times30'$ \citep{2011ExpAstr}; 
the mission duration is also increased from two to five years. 
The image resolutions are $\sim\lambda/D\simeq0".18$ and 83~mas, 
respectively, in a visible spectral band of $\sim 200\, nm$ RMS 
around the central wavelength $\lambda=600\, nm$. 
The expected performance on $\gamma$ scales from $2\times10^{-7}$ for 
the small mission case to $4\times10^{-8}$ for the medium class version. 

\begin{figure}
\begin{centering}
\includegraphics[scale=0.5]{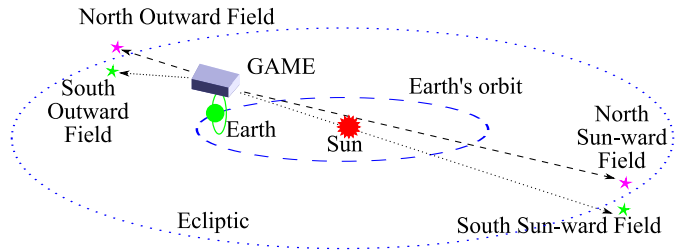}
\par\end{centering}
\caption{\label{fig:fields_1}Representation of GAME observing two Sun-ward 
and two outward fields close to the Ecliptic plane, at $\pm 2^\circ$ latitude. }
\end{figure}

A representation of the GAME operation framework is shown in Fig.~\ref{fig:fields_1}.
The satellite is in a near polar, Sun-synchronous orbit (1500~km
altitude), and observes simultaneously regions either close to the
Sun (hereafter, Sun-ward direction), or opposite to the Sun (outward
direction). 
The fields are superposed in the instrument field of view, using 
techniques similar to those adopted for building a base angle 
in either Hipparcos or Gaia. 
Two fields are placed in simmetric positions above and below the Ecliptic 
plane, and therefore labelled as North and South fields, in both 
Sun-ward and outward direction. 
Ecliptic coordinates will be usually used throughout this paper to best 
match the instrument and operation geometry. 

\begin{figure}
\begin{centering}
\includegraphics[scale=0.6]{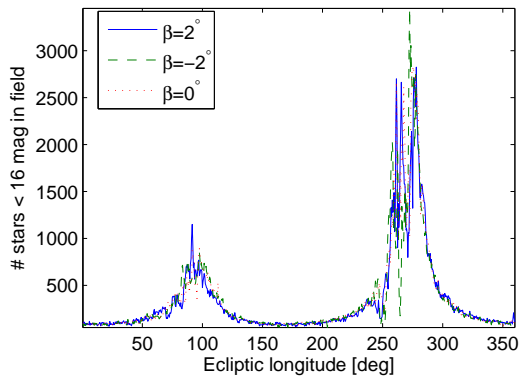}
\par\end{centering}
\caption{\label{fig:StarDensity}Histogram of the number of stars in a 
$30'\times 30'$ field, at latitude 
$\left\{0^\circ, \, \pm 2^\circ \right\}$, 
vs. Ecliptic longitude. }
\end{figure}

\begin{figure}
\begin{centering}
\includegraphics[scale=0.6]{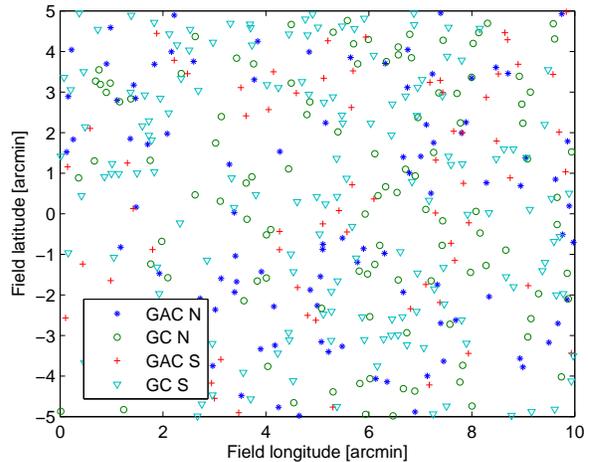}
\par\end{centering}
\caption{\label{fig:FieldCrowding}Sample of superposition of four 
crowded fields in the vicinity of the Galactic centre/anti-centre 
(GC/GAC), at $\pm 2^{\circ}$ Ecliptic latitude, down to 16~mag. }
\end{figure}

The most convenient observing regions, due to the high stellar density, 
are placed at the intersections of the Ecliptic and Galactic planes, 
close to the Galactic Centre and Anti-Centre directions, hereafter 
labelled respectively GC and GAC. 
The limiting magnitude of the GAME stellar sample, compatibly with the 
expected background of residual straylight from the Sun disc and the 
solar corona, is $\sim16$~mag. 
In Fig.~\ref{fig:StarDensity}, the density of isolated stars within 
a $30'\times30'$ field, down to 16~mag, is shown as a function of the 
ecliptic longitude $\lambda$, on the ecliptic plane ($\beta=0^{\circ}$) 
and at the nominal observing latitude ($\beta=\pm2^{\circ}$). 
The complete circular strips ($360^\circ$) contain $\sim2.4\times10^{5}$ 
stars each; the GC and GAC regions, over a $30^{\circ}$ longitude arc, 
include respectively $\sim 9 \times 10^4$ and $\sim 3 \times 10^4$ 
stars, i.e. about 50\% of the total. 
The GC/GAC scan, following the apparent motion of the Sun along the 
Ecliptic, takes about one month. 
 The data are taken from the GSCII catalogue \citep{2008IAUS..248..316B}. 
Crowding affects a marginal fraction of sources, since at the GAME 
limiting magnitude the average star separation in the highest density 
regions is still a few ten arcsec. 
In Fig.~\ref{fig:FieldCrowding}, a typical example of superposition 
of four $10'\times10'$ fields is shown, at $\beta=\pm2^{\circ}$ ecliptic 
latitude and GC/GAC direction ($\lambda=270^{\circ}$ and 
$\lambda=90^{\circ}$ ecliptic longitude, respectively), 
down to 16~mag. 
A small fraction of stars has separation of $\sim1"$, so that they
are still resolved assuming a PSF size of order of 100~mas RMS. 

The BA value, depending on a trade-off between coronagraphic requirements
and deflection amplitude, is $4^{\circ}$. Typically, fields at $2^{\circ}$from
the Sun centre will be observed, which affected by an individual peak
star displacement from its nominal position of $\Delta\psi\simeq230\, mas$,
as shown in Fig.~\ref{fig:PlotDefl}. 

In order to achieve e.g. the $\sigma\left(\gamma\right) \simeq 
\sigma\left(\gamma\right)/\gamma = 4\times10^{-8}$ 
precision goal, associated to the medium mission case, it is then
necessary to measure a few $10^{5}$ stars at the few $\mu as$ level,
exploiting the photon limited precision. 
The requirement on deflection measurement precision can be derived 
from Eq.~\ref{eq:gamma} by error propagation: 
\begin{equation}
\sigma\left(\gamma\right) = \sigma\left(\Delta\psi\right) \cdot 
\frac{c^{2}d}{GM} \sqrt{\frac{1-\cos\psi}{1+\cos\psi}} = 
2 \frac{\sigma\left(\Delta\psi\right)}{\Delta\psi} \,. 
\label{eq:SigmaGamma} 
\end{equation} 
In turn, the deflection angle determination is limited on one side
by \textit{location precision}, i.e. photon noise and instrumental
terms contributing to random errors, and on the other by astrometric
noise on each star. 
Setting the sample size to $N_{S}\approx3\times10^{5}$ stars, measured
over $N_{E}=5$ epochs, the average precision requirement on individual
stars becomes 
\begin{equation}
\sigma\left(\Delta\psi\right) = \frac{1}{2} 
\sigma\left(\gamma\right) \cdot \Delta\psi \sqrt{N_{S}N_{E}} 
= 5.6\,\mu as \,. 
\label{eq:PrecLoc}
\end{equation}
Scaling to the small mission case, the individual precision becomes 
$\sigma\left(\Delta\psi\right) = 28\,\mu as$. 
In both cases, this is compatible with the photon limit of the stellar 
sample for long observations, typically built by composition of shorter 
exposures. 

Observation of different fields implies the need for calibration of 
the independent optical channels with precision adequate to the 
measurement goals. 
Adoption of a combination concept using a large part of the instrument 
in common mode alleviates the implementation constraints, but does not 
modify the requirement value. 
For GAME, in order to retain the residual astrometric errors between nearby 
star images (separation $\sim 10"$) to the $\mu as$ level, the optical scale 
calibration must achieve a precision of order of $10^{-7}$. 

\section{Multiple field astrometry}
\label{sec:MultipleField}
The rationale for superposition of at least two fields is not so much
related to {\em multiplexing efficiency}, since more stars are
observed in the same exposure time, but rather to {\em measurement
accuracy}: the satellite attitude and other instrumental factors
affect in the same way the signal of all targets, allowing their
rejection from differential astrometry as {\em common mode} disturbances.
The observation of many Sun-ward and outward field pairs is achieved
e.g. by sequential pointing of the satellite on a new direction after 
each exposure. 
\\ 
Derivation of angular separation between far away stars, with one-dimensional 
coordinates $\psi_1$ and $\psi_2$, e.g. by tiling 
small field observations from a conventional telescope, would introduce 
at each step the instrument pointing error, which is often larger than 
the typical location precision. 
In such a case, the star angular separation, as well as individual 
deflections, are not well defined from an operational standpoint. 
On the contrary, a pointing correction model is frequently defined, based 
on star observations. 
Pointing errors are, to a high degree, common mode in superposed fields, 
within the precision or stability of the device operating the superposition,
i.e. the {\em beam combiner} (BC). 
An instrument offset is the same for all targets, thus vanishing in 
the focal plane coordinate difference $\Delta \psi_{FP}$ providing 
the star separation on the sky as $\psi_1 - \psi_2 = \Delta \psi_{FP} 
+ \theta_{BA}$. 
The nominal base angle value between the observed fields is 
$\theta_{BA} =4^\circ$; the actual value may be defined by calibration, 
as for Hipparcos, or by an external reference, e.g. metrology 
\citep{2009A&A...507.1739S}. 

\begin{figure}
\begin{centering}
\includegraphics[scale=0.5]{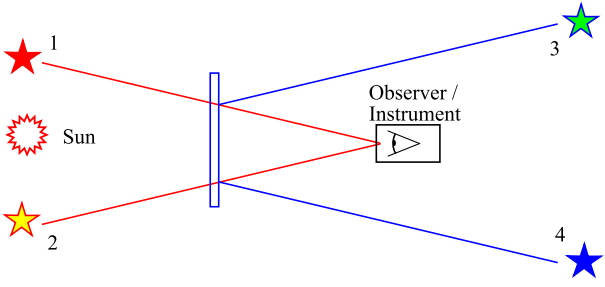}
\par\end{centering}
\caption{\label{fig:MultiFieldEpoch1}Observation in epoch 1: stars 1 and 2 
are at peak deflection value (stars 3 and 4 are at minimum deflection). }
\end{figure}

\begin{figure}
\begin{centering}
\includegraphics[scale=0.5]{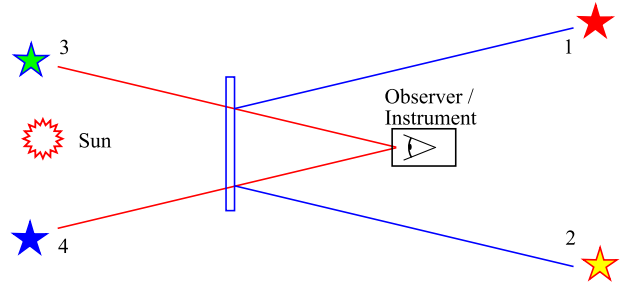}
\par\end{centering}
\caption{\label{fig:MultiFieldEpoch2}Observation in epoch 2: stars 3 and 4 
are at peak deflection value (stars 1 and 2 are at minimum deflection). }
\end{figure}

The differential measurement of deflection on two stars labelled 1
and 2, at symmetric ecliptic latitude $\psi_{1} \simeq \theta_{BA}/2$ 
and $\psi_{2} \simeq -\theta_{BA}/2$, consists in a Sun-ward observation, 
at an epoch 1 (time $t_1$) in which their ecliptic longitude 
$\lambda_1 \simeq \lambda_2$ is close to the Sun centre position, 
plus an observation at an epoch 2 (time $t_2$) separated by six 
months, in which the star direction is opposite to the Sun. 
The former epoch is associated with the maximum value of deflection for 
both stars ($\Delta\psi_{1} = -\Delta\psi_{2}=0".23$); the latter 
corresponds to the minimum deflection value ($\Delta\psi_{1,2} = 
\pm71\,\mu as$), which will be often considered zero throughout the 
discussion for simplicity. 
Comparing the images taken in both epochs (at time $t_{1}$ and $t_{2}$), 
the separation between the stars changes by an angular amount
$\Delta\psi\left(1,2\right) \simeq -\Delta\psi\left(3,4\right)
\simeq 0".46$, 
corresponding to the deflection modulation (adopting the notation:
$\Delta\psi\left(m,n\right)=\Delta\psi_{m}-\Delta\psi_{n}$). 
\\ 
The situation is depicted in Fig.~\ref{fig:MultiFieldEpoch1} and 
\ref{fig:MultiFieldEpoch2}, respectively referred to epochs 1 and 2, 
in which the instrument is always oriented toward the Sun, 
i.e. the deflection HIGH region, and the observation of the opposite 
deflection LOW region is represented by reflection on an ideal flat mirror. 

Notably, the deflection has to be taken into account at practically 
any distance from the Sun, as non negligible at the $\mu as$ level; 
however, in practice, using the nominal deflection value, associated 
to $\gamma=1$, introduces negligible errors on star positions when 
the distance to the Sun is large: e.g., at $45^{\circ}$, the amount 
of deflection is $\left|\Delta\psi\right|=\sim10\, mas$, and even 
a $\gamma$ deviation from unity in the $10^{-5}$ range modifies 
it at the $\sim0.1\,\mu as$ level. 
The $\gamma=1$ assumption is thus fully acceptable, at least as a 
first approximation, in the practical data reduction steps. 
\footnote{Using such estimated location for a new estimate of $\gamma$ would
introduce a propagated error of order of $10^{-7}$per star, i.e.
compatible with the final goal of the mission. Of course, known systematic
errors, as for the above simple approximation, can be taken easily
into account in a realistic data reduction chain, e.g. adopting an
iterative method. 
}

The relative displacement between stars 1 and 2 is twice as large 
as the individual displacement from the nominal position. 
The measured angular separation changes over six months by a comparably 
large amount $\Delta\psi_{1}-\Delta\psi_{2} \simeq 0".466$. 
An asymmetric field placement would provide a significantly smaller 
astrometric signal, e.g., setting one field close to the Sun and the other 
at the ecliptic pole, by a factor $\sim 2$. 
Also, the system symmetry plays an important role throughout the whole 
measurement and calibration process. 

The two epoch observation introduces the fundamental function of 
{\em deflection modulation} on the images of a given field. 
The Sun-ward and outward directions correspond respectively to 
deflection ``HIGH'' and ``LOW'' condition. 

\subsection{Double differential determination of deflection}
\label{sub:Compensation}
The benefits of \textit{simultaneous} observations of deflection ON
and deflection OFF fields are now addressed; in Sec.~\ref{sec:Implementation}
the practical implementation will be dealt with. 
For simplicity, the problem is set as one-dimensional, and 
the instrument is able to perform dual field observations 
by superposing focal plane images from sky regions separated by 
the BA. 
\\ 
Two star, labelled 1 and 2, are selected respectively in the North
and South fields, which are symmetric vs. the ecliptic: 
$\psi_{1} \approx -\psi_{2} \approx \theta_{BA}/2 = 2^{\circ}$, 
where $\psi_{1}$ and $\psi_{2}$ are their ``true'' positions (ecliptic 
latitude) on the sky; the ecliptic longitude corresponds to that of 
the Sun at the observation epoch. 
Similarly, two other stars (3 and 4) are selected in the
opposite direction with respect to the Ecliptic plane ($\pm 180^{\circ}$
in ecliptic longitude). 
The sources are imaged on the focal plane at coordinates $x_{1}$ and 
$x_{2}$ (resp. $x_{3}$ and $x_{4}$), expressed in angular units; taking 
into account the offset $\theta_{BA}$ imposed by the BA, their measured separation is 
$x_{1}-x_{2}=\psi_{1}-\psi_{2}+\theta_{BA}$ 
(resp. $x_{3}-x_{4}=\psi_{3}-\psi_{4}+\theta_{BA}$). 
Each star is affected by a comparable deflection value: 
$\left|\Delta\psi_{n}\right| \simeq 0".23\,,\ n = 1, 2, 3, 4$. 
%
The instrument is switched between the deflection HIGH and LOW fields, 
i.e. star pairs $\{1,2\}$ and $\{3,4\}$, by insertion and removal of 
the flat mirror. 
{\em Simultaneous} observation of the four fields, ideally, can be 
achieved e.g. by replacement of the flat mirror with a semi-reflecting 
mirror, featuring 50\% reflectance and 50\% transmittance. 
The benefit is not related to observation 
efficiency (since the same stars are observed for twice as much time 
with 50\% throughput), but to {\em systematic error control}. 

The equations expressing the star image separation in the two epochs
are now expanded to include explicitly the deflection modulation $\Delta\psi\left(1,2\right)$
(acting in epoch 1), and a possible evolution over time of relevant
instrumental parameters, appearing as a \textit{base angle variation} 
$\Delta\theta_{BA}\left(t_{1},t_{2}\right) = \theta_{BA}\left(t_{1}\right)-\theta_{BA}\left(t_{2}\right)$. 
The measured separation of stars 1 and 2 in the two epochs, 
is respectively 
\begin{equation}
x_{1}\left(t_{1}\right) - x_{2}\left(t_{1}\right) = 
\psi_{1} - \psi_{2} + \theta_{BA}\left(t_{1}\right) + 
\Delta\psi\left(1,2\right)\,;
\label{eq:Defl_ON12}
\end{equation}
\begin{equation}
x_{1}\left(t_{2}\right) - x_{2}\left(t_{2}\right) = 
\psi_{1} - \psi_{2} + \theta_{BA}\left(t_{2}\right)\,, 
\label{eq:Defl_OFF12}
\end{equation}
so that it is possible to subtract of Eq.~\ref{eq:Defl_OFF12} 
from Eq.~\ref{eq:Defl_ON12}: 
\begin{multline}
\Delta\psi\left(1,2\right) + \Delta\theta_{BA}\left(t_{1},t_{2}\right) = \\ 
\left[x_{1}\left(t_{1}\right) - x_{2}\left(t_{1}\right)\right] - 
\left[x_{1}\left(t_{2}\right) - x_{2}\left(t_{2}\right)\right]\,. 
\label{eq:DeflModulRaw12}
\end{multline}
Similarly, the measured separation of stars 3 and 4 is 
\begin{equation}
x_{3}\left(t_{1}\right)-x_{4}\left(t_{1}\right) = 
\psi_{3}-\psi_{4}+\theta_{BA}\left(t_{1}\right)\,;
\label{eq:Defl_OFF34}
\end{equation}
\begin{equation}
x_{3}\left(t_{2}\right) - x_{4}\left(t_{2}\right) = 
\psi_{3} - \psi_{4} + \theta_{BA}\left(t_{2}\right) + 
\Delta\psi\left(3,4\right) \,,
\label{eq:Defl_ON34}
\end{equation}
and by subtracting Eq.~\ref{eq:Defl_OFF34} from Eq.~\ref{eq:Defl_ON34}: 
\begin{multline}
\Delta\psi\left(3,4\right) - \Delta\theta_{BA}\left(t_{1},t_{2}\right) = \\ 
\left[x_{3}\left(t_{2}\right) - x_{4}\left(t_{2}\right)\right] - \left[x_{3}\left(t_{1}\right) - x_{4}\left(t_{1}\right)\right] \,.
\label{eq:DeflModulRaw34}
\end{multline}

The deflection modulation is in phase opposition, and both field pairs 
contribute similarly to the deflection measurement. 
By algebraic composition of Eqs.~\ref{eq:DeflModulRaw12} and 
\ref{eq:DeflModulRaw34}, it is possible to factor out a 
``cumulative deflection'' $\Delta\psi_{1234} = 
\Delta\psi_{1} - \Delta\psi_{2} + \Delta\psi_{3} - \Delta\psi_{4} 
\simeq 0".92 $: 
\begin{equation}
\begin{split}
\Delta\psi_{1234} = & \left[
\Delta x\left(1,2; t_{1}\right) - \Delta x\left(1,2; t_{2}\right) 
\right] + \\ & \left[
\Delta x\left(3,4; t_{2}\right) - \Delta x\left(3,4; t_{1}\right) 
\right] \,; 
\label{eq:DeflMod1234}
\end{split}
\end{equation}
\begin{equation}
\begin{split}
\Delta\theta_{BA}\left(t_{1},t_{2}\right) \simeq & \left[
\Delta x\left(1,2; t_{1}\right) - \Delta x\left(1,2; t_{2}\right) 
\right] - \\ & \left[
\Delta x\left(3,4; t_{2}\right) - \Delta x\left(3,4; t_{1}\right) 
\right] \,; 
\label{eq:dBA1234}
\end{split}
\end{equation}

It may be noted that Eq.~\ref{eq:DeflMod1234} provides a combination 
of deflection values 
{\em corrected, at first order, from the BA variation}. 
Conversely, Eq.~\ref{eq:dBA1234} is an estimate of the BA variation 
{\em deprived, at first order, of the deflection modulation}. 
\\ 
The $\simeq$ sign in Eq.~\ref{eq:dBA1234} appears because the term 
$\delta\Delta\psi = \Delta\psi\left(1,2\right) - 
\Delta\psi\left(3,4\right) \ll 1"$ 
is not considered for simplicity of the expression; in fact, it is
a small but not negligible quantity, depending on the star positions; 
therefore, it is known and can be included in the actual data 
reduction. 

By rearranging the terms, the BA variation corresponds to a variation 
between epochs of a simple quantity corresponding to the ``cumulative 
separation'' of stars in the HIGH and LOW deflection field pairs: 
\begin{equation}
\tilde{x} = \left(x_{1} - x_{2}\right) + \left(x_{3} - x_{4}\right) \,, 
\label{eq:x_ave}
\end{equation}
so that Eq.~\ref{eq:dBA1234} 
(in either compact or complete form) becomes
\begin{equation}
\Delta\theta_{BA}\left(t_{1},t_{2}\right) = \tilde{x}\left(t_{1}\right) - 
\tilde{x}\left(t_{2}\right)\; \left(+\delta\Delta\psi\right)\,.
\label{eq:dBA1234_1}
\end{equation}
Similarly, the cumulative deflection can be rearranged in terms of a 
``differential separation'' of stars: 
\begin{equation}
\hat{x} = \left(x_{1} - x_{2}\right) - \left(x_{3} - x_{4}\right) \,, 
\label{eq:x_diff}
\end{equation}
so that Eq.~\ref{eq:DeflMod1234} becomes 
\begin{equation}
\Delta\psi_{1234} = 
\hat{x}\left(t_{1}\right) - \hat{x}\left(t_{2}\right) \,.
\label{eq:DeflMod1234_1}
\end{equation}

The proposed combination of measurements bears the promise of a simple,
effective and ``clean'' deflection determination, leading to a
robust estimate of the PPN $\gamma$ parameter. 
The actual performance over a given star sample (with the above equations 
extended to include the coordinates of all targets) depends on the actual 
source brightness distribution, as high precision observations contribute 
to both deflection estimate (Eq.~\ref{eq:DeflMod1234}) and BA monitoring 
(Eq.~\ref{eq:dBA1234}) in the same way. 
\begin{figure}
\begin{centering}
\includegraphics[scale=0.45]{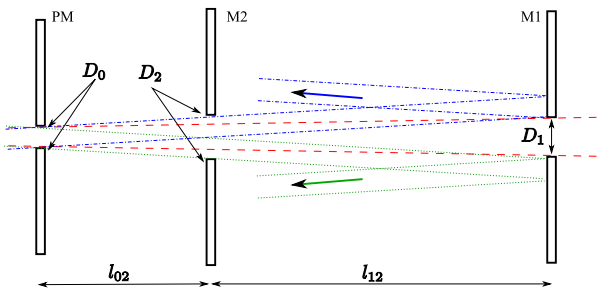}
\par\end{centering}
\caption{\label{fig:Layout1}Schematic of the individual aperture beam
path: the Sun beam (dashed lines) from the pupil mask PM gets out of 
the aperture on the primary mirror M1, whereas the  stellar beams 
(dotted and dash-dot lines) are captured by M1. }
\end{figure}

\section{Optical implementation concept}
\label{sec:Implementation}
The GAME optical design is based on Fizeau interferometry, in order
to achieve a convenient trade-off between the angular resolution needed
for precision astrometry, and coronagraphic requirements, applied
to small apertures achieved by pupil masking on the underlying telescope.
The beam path related to a single input aperture is described below,
in a numerical example related to the small mission version; then,
the multiple aperture combination is considered. 
\\ 
The elementary aperture is circular, with diameter $D_{0}=4\, cm$. 
A schematic view of the basic layout is shown in Fig.~\ref{fig:Layout1}:
the input aperture on the pupil mask PM feeds the instrument with
beams of diameter $D_{0}$; the beams from the Sun (dashed lines)
and stars from North and South Sun-ward fields (dotted and dash-dot
lines, respectively) are separated in terms of geometric optics on
the first mirror M1, at a distance $l_{01}=l_{02}+l_{12}=1.5\, m$ 
from PM, where an output aperture of suitable diameter $D_{1}>D_{0}$ 
outputs the solar photons toward outer space. 
The output aperture must be larger than the input one due to 
(a) the finite angular size of the Sun ($R_{\odot}\simeq16'$); 
and (b) the margins with respect to the geometric shadow 
edge. 
The apodisation design is 
discussed in \cite{2010SPIE.7731E..53L}. 
The current value is $D_{1}=6\, cm$. 

M1 represents the primary mirror of the telescope; M2 is the flat 
folding mirror used to feed the outward field beams into the instrument, 
as required in Sec.~\ref{sec:MultipleField}. 
M2 does not affect the Sun-ward beams, passing through suitable apertures 
($D_2>D_0$ to account for BA and field size). 
The distances from M2 to PM and M1 are respectively $l_{02}=0.45\, m$
and $l_{12}=1.05\, m$. 
The beams from the North and South Sun-ward fields (direction 
$\pm\theta_{BA}/2$) are totally collected by the primary 
mirror, under the geometric constraint that the stellar beam edges 
(position $\pm D_{0}/2$ on PM) falls outside the opposite edge of the 
output aperture (position $\mp D_{1}/2$ on M1). 
Therefore, 
\begin{equation}
l_{01} \cdot \frac{\theta_{BA}}{2} \geq \frac{D_0+D_1}{2} \, , 
\label{eq:GeomConstr1}
\end{equation}
neglecting the beam divergence due to the finite field size.
\begin{figure}
\begin{centering}
\includegraphics[scale=0.45]{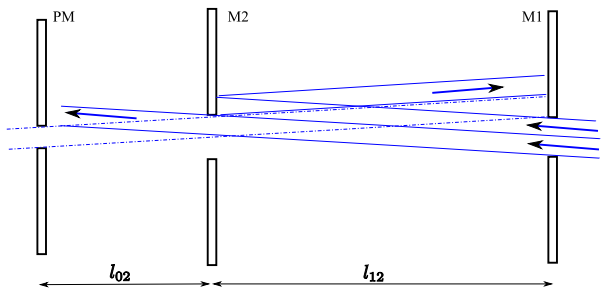}
\par\end{centering}
\caption{\label{fig:Layout2}Schematic for the outward field beams: 
the outward South beam (solid line), back-reflected by M2, is injected 
onto M1, parallel to the Sun-ward South beam (dash-dot lines), 
partially vignetted by both mirrors. } 
\end{figure}

The mirror M2 is used to inject the beams from the outward fields 
into the instrument, as represented in Fig.~\ref{fig:Layout2}. 
The aperture on M2, with radius 
$R_{2}\geq R_{0}+l_{02} \cdot \theta_{BA}/2=R_{0}+1.5\, cm$, 
allows unhindered transit to the stellar beams. 
\begin{figure}
\begin{centering}
\includegraphics[scale=0.75]{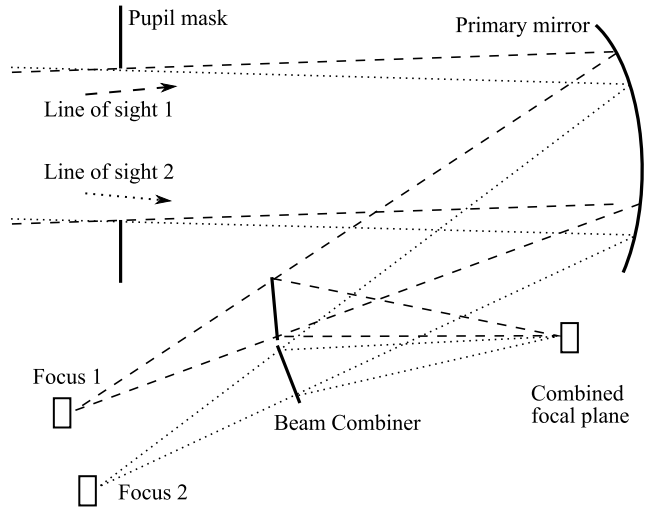}
\par\end{centering}
\caption{\label{fig:BeamComb}Folding of the beams from fields of view 
1 and 2 (dashed and dotted lines) onto a common focal plane by means 
of a Hipparcos-like beam combiner. } 
\end{figure}
The beam from the South outward direction, shown in Fig.~\ref{fig:Layout2}, 
is partially vignetted by the M2 aperture, since it is displaced by 
$l_{12} \cdot \theta_{BA} = 3.5\, cm$. 
Part of the South outward beam is reflected back by M2 towards M1, parallel 
to the South Sun-ward beam from PM, thus effectively superposing on the 
telescope focal plane the images of both front and rear fields. 
By symmetry, injection onto M1 by reflection on M2 is achieved also to 
the North Sun-ward and outward beams. 
With proper geometry, the beam size can be made comparable for the front 
and rear viewing direction, ensuring similar photon throughput. 

The further step in the system definition is the Fizeau combination of 
several beams and apertures to achieve the desired angular resolution, 
associated to an underlying larger telescope. 
The aperture separation on PM must be such to accommodate on M1 the
Sun beam output and the footprints of North and South stellar beams,
plus a suitable margin to accommodate the back reflected outward beams 
with acceptable vignetting. 

The concept of simultaneous observation on four field described in 
Sec.~\ref{sec:MultipleField} is therefore implemented by {\em wavefront 
division}, rather than {\em amplitude division}. 
Since usage of small apertures is imposed by the need of separating solar 
and stellar photons, the separation among apertures is taken advantage 
of to achieve the desired multiplexing of observing directions. 
With respect to the simple concept of a semi-reflecting mirror, it may 
be noted that in this case no transmitting element is introduced on the 
beam paths; this might have introduced astrometric errors due to 
material inhomogeneity, and increased straylight from scattering. 

In this way, {\em the desired four field observing instrument, pair-wise 
symmetric} and using as far as possible the same components in common 
mode, is achieved. 
The penalty for injection of the outward beams is the additional mirror 
M2 (still less cumbersome than a whole duplicated telescope) and $\sim10\%$ 
increase of the aperture array spacing, from the minimum value of 
$D_0 + D_1 = 10\, cm$ to $11\, cm$. 
The optical engineering aspects of the telescope, after M1, are not 
further detailed herein. 
A conceptual representation of the beam folding onto a common focal 
plane, e.g. by a Hipparcos-like beam combiner, is shown in 
Fig.~\ref{fig:BeamComb}. 

\begin{figure}
\begin{centering}
\includegraphics[clip,trim=0mm 10mm 0mm 15mm,scale=0.5]{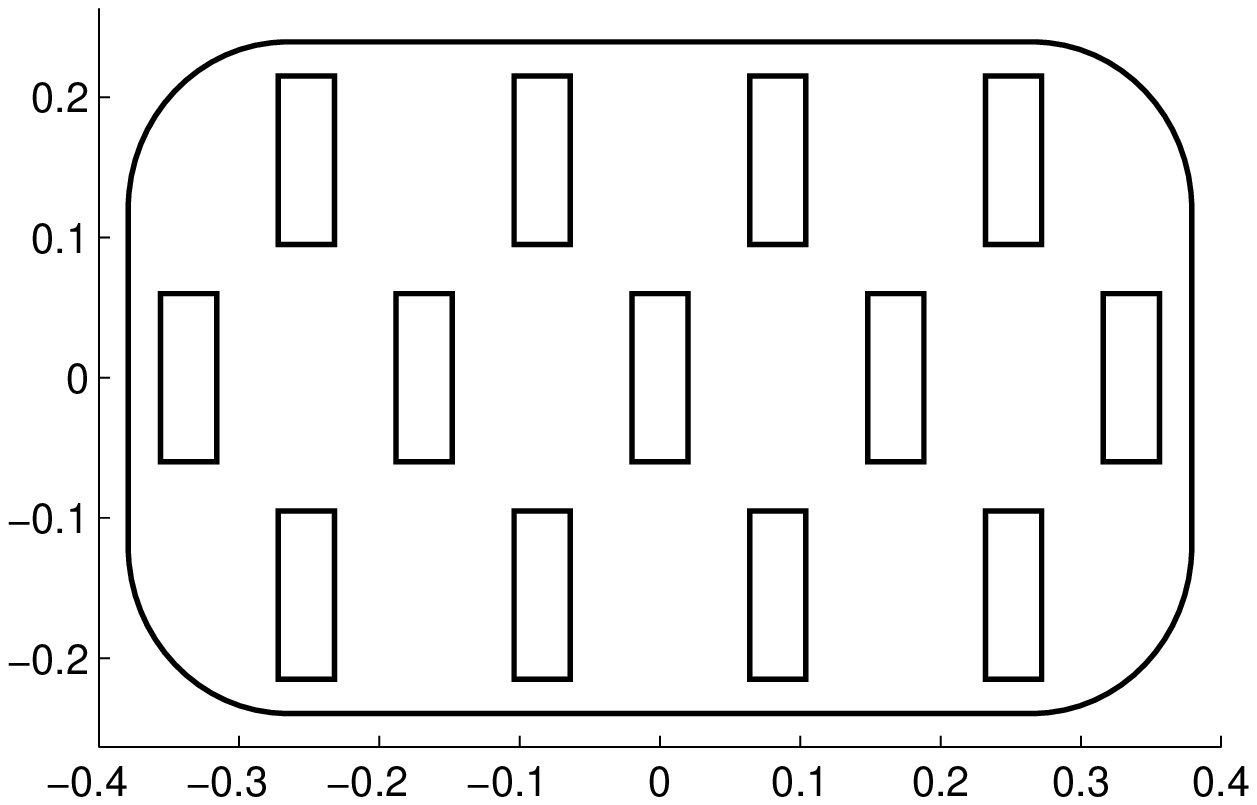}
\includegraphics[clip,trim=0mm 10mm 0mm 10mm,scale=0.5]{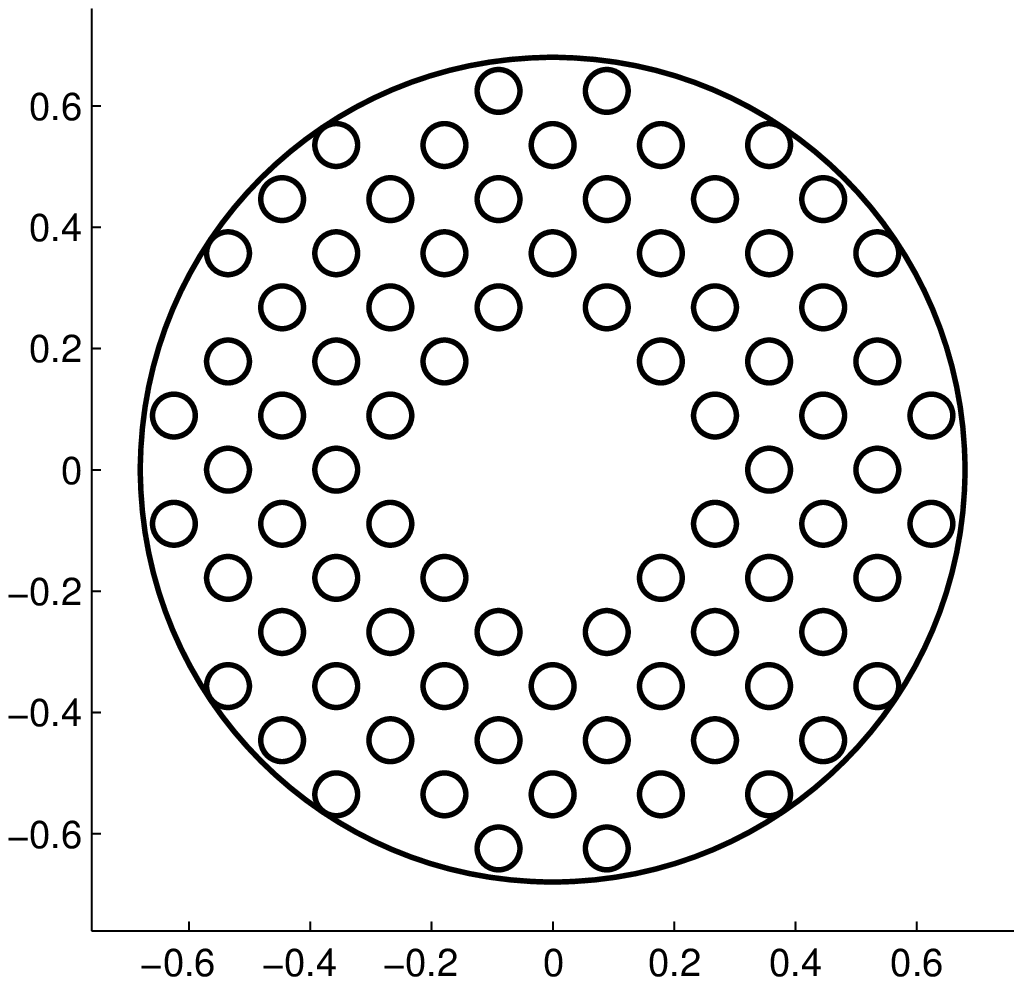}
\par\end{centering}
\caption{\label{fig:PupilMask}Pupil mask for the small mission version (top) 
and the medium mission version (bottom) of GAME. }
\end{figure}

A set of 13 elementary apertures fitting the above geometry and the
overall size constraint $\sim 0.5\times0.7\, m$ is shown in Fig.~\ref{fig:PupilMask} (top).
The geometry is suited to an off-axis telescope, and considered compatible
with the envelope of a small mission \citep{2009SPIE.7438E..20G}.
Similarly, the pupil mask defined for the envelope of a medium class 
mission \citep{2011ExpAstr}, optimised for a centred telescope, includes 
80 elementary apertures and it is shown in Fig.~\ref{fig:PupilMask} 
(bottom).

A filled aperture telescope with resolution corresponding to the equivalent
$0.8\, m$ beam of the small mission version would require a distance
to the occulter of $23\, m$ to achieve the geometric optics separation
of the beams from the Sun and the fields; the distance increases to
$35\, m$ for the $1.5\, m$ medium mission version. 
This obviously would not fit the size of a conventional satellite,
therefore requiring e.g. formation flying solutions. The proposed
Fizeau solution achieves a comparable angular resolution in a much
more compact envelope, suitable to a practical payload allocation.

\subsection{Differential instrument response}
Due to the high angular precision goal, at the micro-arcsec level, 
the small unavoidable differences among fields of view are potentially 
relevant, and they have to be adequately (inter-)calibrated, as 
discussed in Sec.~\ref{sec:Calibration}. 
However, it may be of interest to briefly address here some of the 
relevant instrumental aspects of the proposed solution. 

The two Sun-ward fields (and, similarly, the two outward fields) 
correspond to symmetric regions of the first focal plane of the 
initial section of the telescope, so that they have symmetric 
response in the nominal design. 
Tolerances and alignment will degrade the symmetry, but, with 
proper implementation care, strict correlations may be retained; 
considerations are presented e.g. in \cite{2006A&A...449..827B}. 
The adoption of a beam combiner, superposing the fields on the final 
focal plane, partially alleviates the matter because the common parts 
of the system are maximised. 

A clear difference between the Sun-ward and outward field pairs is 
the additional reflection of the latter beams on mirror M2. 
In the limiting case of an ideal flat mirror, its displacement 
(piston and decenter) does not introduce significant astrometric errors 
but only a marginal beam vignetting. 
The M2 mirror tilts induce the same deviation on {\em both} outward 
beams, so that no astrometric error is inserted in the measurement. 
The mutual displacement of the outward field image pairs 
with respect to the Sun-ward ones is immaterial, as it does not 
contribute to deflection determination (Sec.~\ref{sub:Compensation}). 

Deviations of M2 from the nominal flatness introduce an additional 
wavefront error on the outward fields, modifying the optical response, 
which at first order can be represented by the optical scale, associated 
to the mirror curvature. 
The implementation of scale calibration is discussed in 
Sec.~\ref{sec:Calibration}, neglecting at the moment the issues 
related to image profile variation, which can be addressed e.g. as 
in \cite{2010MNRAS.406.2433G}. 

The expected instrument asymmetry can be appreciated through a simple model 
of thin lens combination. 
If $f_1$ is the telescope focal length associated to the Sun-ward fields, 
the focal length $f_2 = R_2 / 2$ associated to a finite curvature radius 
$R_2$ on M2 induces an effective focal length $f_e$ on the outward 
fields such that 
\begin{equation}
\frac{1}{f_e} = \frac{1}{f_1} + \frac{1}{f_2} - \frac{l_{12}}{f_1 f_2} 
\simeq \frac{1}{f_1} + \frac{1}{f_2} 
\, , 
\label{eq:EFL}
\end{equation}
using the thin lens approximate expression, where $l_{12}$ is the distance 
between mirrors M1 and M2. 
The M2 curvature radius associated to a sagitta comparable to the visible 
wavelength between mirror centre and edges, $\delta \sim 100\, nm$, i.e. 
a significant error for a good optical quality component, is 
\begin{equation}
R_2 = \frac{r_{M2}^2 + \delta^2}{2 \delta} \sim 10^6 \, m 
\, , 
\label{eq:R2}
\end{equation}
where $r_{M2} \sim 0.5\, m$ is the semi-chord length (mirror radius). 
The focal length of M2 is then $f_2 \sim 5 \times 10^5 \, m$. 
The optical scale is the reciprocal of the effective focal length, i.e. 
$s_1 = 1/f_1$ for the Sun-ward fields, and $s_e = 1/f_e$ (Eq.~\ref{eq:EFL}) 
for the outward fields. 
The relative variation of optical scale is 
\begin{equation}
\frac{\Delta s_e}{s_e} =
\frac{f_1}{f_1 + f_2} \sim 5 \times 10^{-5} \, m 
\, . 
\label{eq:Delta_s}
\end{equation}
Such variation corresponds to an apparent displacement of 
$50\, \mu as$ for two stars separated by a constant $10"$ angle, when 
observed in either the Sun-ward or outward fields. 

\section{Optical scale calibration}
\label{sec:Calibration}
The evolution of instrument parameters is potentially critical to 
high precision measurements. 
It is thus crucial to provide an indication of convenient methods 
to ensure that the system characteristics are monitored with adequate 
precision. 
Since the star separations are directly inserted into the deflection
measurement (Eqs.~\ref{eq:Defl_ON12} to \ref{eq:dBA1234_1}), the
uncorrected scale variation enters as a {\em systematic error},
appearing as common mode over all target combinations, and are 
therefore not averaged over the number of objects in the field. 
The instrument response variation {\em might} average down throughout 
the mission, but this cannot be taken for granted, and the trend 
might be dangerously slow. 
The approach described below, focused on the optical scale $s$, takes 
advantage of repeated observation of fields populated by many stars. 

The optical scale is respectively $s=8.3"/mm$ for the small mission 
version of GAME ($EFL=25\, m$) and $s=5.9"/mm$ for the medium mission 
case ($EFL=35\, m$). 
An image detected in the focal plane position $x$ is associated to a 
source in the sky having angular position $\psi=s\cdot x$ (gnomonic 
projection); a scale variation modifies the image location by an 
amount proportional to the source position, thus introducing an 
error also on the {\em estimated separation} between stars. 
For the current computation, the stars are considered as having constant 
angular position, flux etc., and only instrument parameter variations are introduced. 

The case of one-dimensional measurement of $N$ isolated, well behaved stars 
with given source parameters (magnitude, spectrum) is considered; their 
positions $x_{n}$ ($n=1,2,\ldots,N$) can be estimated with photon limited precision $\sigma_{n}$ \citep{1998PASP..110..848G}: 
\begin{equation}
\sigma_{n} = \alpha \frac{\lambda}{X} \cdot \frac{1}{SNR_{n}} 
\,,
\label{eq:ElemPrec}
\end{equation}
depending on the instrument characteristics (RMS aperture size $X$ and 
observing wavelength $\lambda$), through the photometric 
signal to noise ratio (SNR). 

The {\em field centre} (FC) of the set of stars is defined in the 
following as the average position $x_{0}$, weighted by the individual 
location variance, using the maximum likelihood estimator: 
\begin{equation}
x_0 = \frac{\sum x_n/\sigma_n^2} {\sum1/\sigma_n^2} = 
\sigma_0^2 \cdot \sum \frac{x_n}{\sigma_n^{2}}
\,,
\label{eq:CPC1}
\end{equation}
and the FC variance $\sigma_0$ is 
\begin{equation}
\sigma_0^2 = \frac{1}{\sum 1/\sigma_n^2}\,.
\label{eq:CPC_RMS1}
\end{equation}

In the photon limited case, the individual location error from Eq.~\ref{eq:ElemPrec}
for an observation collecting $p_n$ photons, is $\sigma_n = 
\sigma_I/\sqrt{p_n}$, where the parameter $\sigma_{I}$, used to factor 
out the other relevant parameters, represents formally the instrument 
performance for an object with $SNR = 1$. 
The FC variance becomes $\sigma_0^2=\sigma_I/\sum p_n$, i.e. it is 
naturally related to the total photometric level $P_t = \sum p_n$. 

The {\em field size} (FS) is defined as the RMS distance $x_{RMS}$ 
of each star from the FC, i.e. the quantities $\{x_n-x_0\}$, 
again weighted by the individual location variance: 
\begin{equation}
x_{RMS}^2 = \sigma_0^2 \cdot \sum \frac{\left(x_n-x_0\right)^2}{\sigma_n^2} = 
\sigma_0^2 \cdot \sum \frac{x_n^2}{\sigma_n^2}-x_0^2
\,,
\label{eq:FSize1}
\end{equation}

By construction, the relative distances have zero weighted average. 
However, the RMS distance is a convenient parameter for instrument 
calibration purposes, since it is a differential quantity, independent 
e.g. from pointing. 
Its variance is a function of the FC variance (location precision) and 
the field geometry: 
\begin{equation}
\sigma^2(x_{RMS}) = \sigma_0^2 \left(1+2\frac{x_0^2}{x_{RMS}^2}\right)
\,.
\label{eq:RMS_FS}
\end{equation}
Assuming that the star distribution is approximately uniform 
over the region $\left[-\Omega/2,\Omega/2\right]$, and that their 
location errors are comparable, i.e. $\sigma_n \simeq \sigma$, 
e.g. selecting sources in a limited magnitude and spectral range, 
then
$x_{RMS} \simeq \Omega/\sqrt{12}$, 
$\left\langle x_0\right\rangle \simeq 0$
and 
$\left\langle x_0^2\right\rangle \simeq x_{RMS}^2/N$. 
Therefore, for $N\gg1$, 
\begin{equation}
\sigma^2(x_{RMS}) \simeq \sigma_0^2 \left(1+\frac{2}{N} \right) \approx 
\sigma_0^2 \simeq \frac{\sigma^2}{N}
\,,
\label{eq:RMS_FS_2}
\end{equation}
i.e. {\em both FS and FC have comparable precision}, improving with the 
sample size $N$. 

As a numerical example, applicable to a set of a few elementary exposures 
on a single CCD of the GAME focal plane, with $N\simeq 100$ stars 
measured at $\sigma \simeq 1\, mas$, for a field amplitude 
$\Omega \simeq 4\, arcmin$, the FS is $x_{RMS} \simeq 1'.15$, 
and the FC standard deviation is
$\sigma_0 \simeq \sigma(x_{RMS}) \simeq 100\,\mu as$. 

Assuming a small optical scale variation between epochs $t_1$ and 
$t_2$, so that 
$s\left(t_2\right) = \phi \cdot s\left(t_1\right),\: \phi \simeq 1$, 
each star location and the FC are displaced by an amount proportional 
to their value, i.e. 
$x_n \left(t_2\right) = \phi \cdot x_n \left(t_1\right)$, 
$x_0 \left(t_2\right) = \phi \cdot x_0 \left(t_1\right)$, 
and also 
$x_{RMS} \left( t_2\right) = \phi \cdot x_{RMS} \left(t_1\right)$. 
The variation can be monitored through the geometry of 
repeatedly measured fields: 
\begin{equation}
\phi = \frac{s\left(t_2\right)}{s\left(t_1\right)} = 
\frac{x_{RMS}\left(t_2\right)}{x_{RMS}\left(t_1\right)}
\,,
\label{eq:ScaleMon}
\end{equation}
and is affected by a propagated error 
\begin{equation}
\frac{\sigma(\phi)}{\phi} \simeq \sigma(\phi) \simeq 
2\frac{\sigma(x_{RMS})}{x_{RMS}} \simeq 
\frac{2}{\sqrt{N}} \cdot \frac{\sigma}{x_{RMS}} 
\,.
\label{eq:ScaleErr}
\end{equation}
The precision is high because of the ratio between the 
individual location precision $\sigma$ and the FS $x_{RMS}$. 
It is therefore possible to define a convenient calibration field as 
having a widely spread distribution of comparably bright stars. 
Using the numbers from the above numerical example, the scale 
calibration precision is $\sigma(\phi) \simeq 3 \times 10^{-6}$. 

The optical scale monitoring over a few hours of observations, at 
full focal plane level, applied to the small mission version, with 
$\sigma = 1\, mas$, $\Omega = 14'$, $x_{RMS} \simeq 4'$, $N \simeq 600$, 
achieves a precision $\sigma(\phi) \simeq 3 \times 10^{-7}$; 
similarly, for the medium mission version, with $\sigma = 0.5\, mas$, 
$\Omega = 30'$, $x_{RMS} \simeq 8'.7$, $N \simeq 2,500$, the value 
attained is $\sigma(\phi) \simeq 4 \times 10^{-8}$. 
In both cases, the short term monitoring precision compares 
favourably with the requirements. 

Similar considerations may be applied e.g. to the geometric calibration 
of the two Gaia channels, observing in Time Delay Integration long strips 
of the sky, with significant superposition of their $0^{\circ}.7$ width, 
at a rate of $\sim 60"/s$. 
The elementary exposure precision for medium to bright stars is 
$\sigma < 1\, mas$, so that, on a limited region with $N\simeq 1000$ 
stars, and size roughly corresponding to the astrometric focal plane, 
the optical scale monitoring performance on a data segment 
is $\sigma(\phi) \simeq 10^{-7}$. 
The performance significantly improves on longer strips, 
under the assumption of stability on short to intermediate 
time scales. 
This simple exercise supports the high expectations on Gaia 
self-calibration properties. 

\section{Sample astrometry }
\label{sec:astr_constr}
The general requirement on star location precision (a few $\mu as$) 
is mentioned in the Introduction. 
However, stars are characterised by individual proper motion and 
parallax, modifying their position with time. 
To achieve a given precision level on the $\gamma$ estimate, 
specific requirements on the knowledge 
of the star parameters can be set; some of them, detailed below, 
may be subject to averaging depending on the number of stars 
$N_S$ contributing to the estimate, and/or the number of epoch pairs 
$N_E$ (i.e. years) in which the measurement is repeated. 
E.g. for the medium mission version of GAME, the goal precision is 
$\sigma\left(\gamma\right) \simeq 10^{-8}$ and a sample of 
$N_S \simeq 4 \times 10^5$ stars is observed over $N_E =5$ years. 

The precision requirement on the individual star position $\psi$ 
can be derived by error propagation from Eq.~\ref{eq:gamma}: 
\begin{equation}
\sigma_{\psi}\left(\gamma\right) = 
\frac{\partial\gamma}{\partial\psi} \sigma\left(\psi\right) = 
\frac{1+\gamma}{\sin\psi} \sigma\left(\psi\right)\simeq 
\frac{2}{\sin\psi} \sigma\left(\psi\right) \,.
\label{eq:ErrPos}
\end{equation}
This must be scaled according to the sample size, i.e. by a factor 
$\sqrt{N_S}$. 
Measuring repeatedly the same stars, no averaging on epochs is 
possible, since their individual coordinate error is applied 
each time. 
Then the requirement on individual position uncertainty is 
\begin{equation}
\sigma\left(\psi\right) = 
\frac{\sin\psi}{1+\gamma} \sigma_{\psi} \left(\gamma\right) \sqrt{N_{S}} 
\simeq 18\, mas
\,,
\label{eq:PrecReqPos}
\end{equation}
i.e. challenging for current ground based catalogues (GSCII), but quite 
relaxed with respect e.g. to the expected performance of the forthcoming 
Gaia catalogue in the GAME magnitude range. 

\textit{Proper motion}, in a determination of the light deflection 
using just the two observations taken over one year of operation, 
would introduce a significant error on the deflection estimate, 
if neglected. 
The total star position variation $\Delta\psi_{tot}$ between 
Sun-ward and outward observation epochs, separated by six months, 
is the sum of the apparent angular displacement associated to 
light deflection, i.e. $\Delta\psi$ from Eq.~\ref{eq:Misner}, and 
half the yearly proper motion $\mu$: 
$\Delta\psi_{tot} = \Delta\psi + \frac{1}{2}\mu$. 
\\ 
However, most stars are observed $N_{E}$ times throughout the mission 
lifetime, and the sequence of measured positions includes both 
deflection modulation (with a one year period) and the constant drift 
associated to proper motion. 
The two contributions can easily be separated, 
with precision improving with $N_E^{-3/2}$ 
\citep{2009AJ....137.4400L}. 

\begin{figure}
\begin{centering}
\includegraphics[clip,scale=0.8]{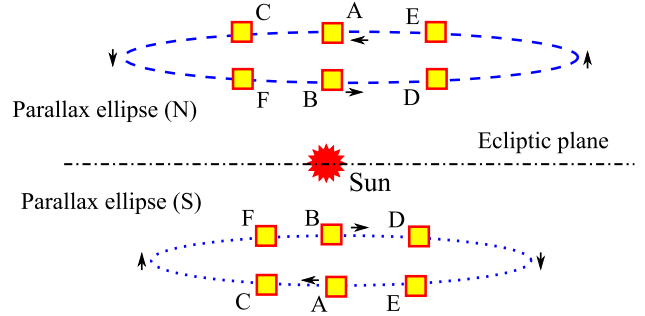}
\par\end{centering}
\caption{\label{fig:ObsStrat}Observing strategy: epochs A and B are the 
nominal deflection determination positions; C and D (and similarly E and F) 
are the parallax determination positions. } 
\end{figure}

\textit{Parallax}, conversely, is most critical, since
the star motion induced by the orbital motion of the Earth around
the Sun has the same period and phase of deflection modulation. 
A schematic of the parallax ellipse, followed in opposite direction 
in the two hemispheres, is shown in Fig.~\ref{fig:ObsStrat}. 
The two measurement epochs of GAME, approximately corresponding 
to the conjunction and opposition of each star with the Sun, 
respectively for high and low deflection conditions, are 
labelled as points $A$ and $B$ in figure. 
At $\beta=\pm2^\circ$, the ellipse is strongly elongated in the 
East-West direction, so that the parallax component in the 
deflection measurement direction (North-South) is reduced by a 
geometric factor $\sim 2\times 2^\circ/90^\circ=0.044$, i.e., for 
a star at $1\, kpc$, to $44\, \mu as$. 

Proper motion and parallax correction at the few $\mu as$ 
level corresponding to the GAME goal appears to be 
marginally compatible with the GAIA catalogue; however, 
it is convenient to define a self-consistent observing strategy
for GAME. 
This can be done by adding further observing epochs, labelled C, 
D, E and F in Fig.~\ref{fig:ObsStrat}, thus allowing full astrometric 
reconstruction for the whole stellar sample, and a most important 
cross-calibration tool. 

Given the proposed concept of multiple field observation, the 
Sun-ward and outward fields are pointed simultaneously in e.g. 
$A$ and $B$ positions (HIGH and LOW deflection) respectively. 
At a time difference of about one month, the fields can be 
observed again at $\pm30^{\circ}$ from the Sun, in the position 
pairs $\left\{ C,\,D \right\}$ or $\left\{ E,\,F \right\}$, 
in different, known phases with respect to both deflection and 
parallax modulation. 
Stars in the $\left\{ C,\,D \right\}$ (and respectively 
$\left\{ E,\,F \right\}$) positions are affected by opposite 
displacement in the East-West direction, due to parallax geometry, 
$\sim 15$ times larger than that suffered in epochs A/B (in the 
North-South direction). 

The repeated observation scheme requires that the GC/GAC regions 
are scanned subsequently three times, i.e. before, during and after 
the Sun conjunction. 
The sequence takes about six months a year, which leaves $\sim 50\%$ 
of the observing time for other science goals. 

The GAME data set from the multiple epoch observing strategy will 
thus provide a relative astrometry catalogue for general purpose 
astronomy, at a precision level comparing well with that of the 
Gaia catalogue, apart serving the needs for calibration and 
self-consistency of the GAME data reduction. 

\section*{Conclusions \label{sec:Conclusions}}

The superposition in a suitable configuration of several fields provides
constraints among the measurements which may allow a significant mitigation
of systematic errors, in particular related to base angle variations. 

The additional benefit is a clear assessment of the measurement 
systematic error related to the base angle from the science data, 
due to the double differential technique, providing both 
self-calibration and monitoring. 

A simple implementation scheme is outlined, which can be easily adapted
to different instrument scales, depending on the performance goal
and other constraints. Due to the flexibility of pupil mask and telescope
geometry of a Fizeau interferometer, the actual design can be tailored
to fit the system requirements, e.g. accommodation of other payloads. 

The considerations on systematic error control from 
Sec.~\ref{sec:MultipleField} must take into account the limitations 
imposed by the real sky distribution of stars: in particular, the terms 
in Eqs.~\ref{eq:DeflMod1234} and \ref{eq:dBA1234} will not have exactly 
the same statistical weight, depending at least on the photon limit. 
In particular, the GC region, due to the higher star count, dominates 
the statistics, with a corresponding precision two to three times better 
than the GAC region; therefore, the systematic error assessment is 
basically limited by the latter, whereas the former sets the limiting 
noise (random error) on the $\gamma$ measurement. 

The GAME concept appears therefore suited to photon limited estimation 
of the PPN parameter $\gamma$, down to the $10^{-7}\, - \, 10^{-8}$ 
level, according to the experiment implementation scale. 

\section*{Acknowledgments}

The study presented in this paper benefits from discussions with 
A. Vecchiato, M. Lattanzi, S. Capozziello, A. Nobili and other 
colleagues. 
The activity was partially supported by the Italian Space Agency 
through contracts ASI I/016/07/0 (COFIS) and ASI I/037/08/0. 

\bibliographystyle{mn2e} 


\label{lastpage}
\end{document}